\documentclass[12pt,preprint]{aastex}

\usepackage{natbib}

\newcommand{\degree}{$^{\circ}$~}
\newcommand{\degreee}{$^{\circ}$}
\newcommand{\kmsec}{km~s$^{-1}$~}
\newcommand{\kmsecc}{km~s$^{-1}$}

\newcommand{\VEX}{\emph{VEX} }

\newcommand{\WIN}{\emph{Wind} }

\begin{document}
\title{Prediction of Geomagnetic Storm Strength from Inner Heliospheric In Situ Observations}
\author{M. Kubicka\altaffilmark{1}, C. M\"ostl\altaffilmark{1,2}, T. Amerstorfer\altaffilmark{1}, P. D. Boakes\altaffilmark{1,2}, L. Feng\altaffilmark{3}, J. P. Eastwood\altaffilmark{4}, O. T\"orm\"anen\altaffilmark{1,5}}

\affil{\altaffilmark{1}Space Research Institute, Austrian Academy of Sciences, 8042 Graz, Austria}
\affil{\altaffilmark{2}Kanzelh\"ohe Observatory-IGAM, Institute of Physics, University of Graz, 8010 Graz, Austria}
\affil{\altaffilmark{3}Purple Mountain Observatory, Chinese Academy of Sciences. West Beijing Road 2 Nanjing, 210008, China}
\affil{\altaffilmark{4}Space and Atmospheric Physics, The Blackett Laboratory, Imperial College London, London SW7 2AZ, UK}
\affil{\altaffilmark{5}Aalto University, School of Electrical Engineering, 02150 Espoo, Finland}

\email{christian.moestl@oeaw.ac.at}

\begin{abstract}
Prediction of the effects of coronal mass ejections (CMEs) on Earth strongly depends on knowledge of the interplanetary magnetic field southward component, $B_z$. Predicting the strength and duration of $B_z$ inside a CME with sufficient accuracy is currently impossible, which forms the so-called $B_z$ problem. Here, we provide a proof-of-concept of a new method for predicting the CME arrival time, speed, $B_z$ and the resulting \textit{Dst} index at Earth based only on magnetic field data, measured in situ in the inner heliosphere ($< 1$~AU). On 2012 June 12--16, three approximately Earthward-directed and interacting CMEs were observed the by the STEREO imagers, and by Venus Express (VEX) in situ at 0.72 AU, 6 degree away from the Sun Earth line. The CME kinematics are calculated using the drag--based and WSA--Enlil models, constrained by the arrival time at \textit{VEX}, resulting in the CME arrival time and speed at Earth. The CME magnetic field strength is scaled with a power law from \textit{VEX} to \textit{Wind}. Our investigation shows promising results for the \textit{Dst} forecast (predicted: $-96$ and $-114$~nT (from 2 Dst models), observed: $-$71 nT), for the arrival speed (predicted: $531 \pm 23$~\kmsecc, observed: $488 \pm 30$~\kmsecc) and timing ($6 \pm 1$ hours after actual arrival time). The prediction lead time is 21 hours. The method may be applied to vector magnetic field data from a spacecraft at an artificial Lagrange point between the Sun and Earth, or to data taken by any spacecraft temporarily crossing the Sun--Earth line. 
\end{abstract}


\section{Introduction}
Extreme space weather events are mostly driven by coronal mass ejections (CMEs) and their interplanetary manifestations, called interplanetary CMEs or ICMEs \citep[e.g.][]{zha07}. They have potential to be hazardous for an extensive list of different technologies, therefore the precise forecasting of such space weather events becomes increasingly important. Many different methods exist to predict the initial speeds, directions and shapes of CMEs  and the evolution of CMEs through the inner heliosphere, but none of them are able to accurately predict the magnetic field components, in particular the southward $B_z$ component, which is crucial to calculate a CME's geomagnetic effectiveness \citep[e.g.][]{dun61,ros67,bur75,hoe92,obr00}.

Here, we present a proof-of-concept for predicting the strength of geomagnetic storms caused by CMEs, using in situ data from a spacecraft situated in the inner heliosphere, i.e.\ in the space $< 1$ AU, and close to the Sun--Earth line. By definition \citep{rou11} we use the term ICME (interplanetary CME) for the in situ observation of a CME. With measurements from \textit{Venus Express} (\textit{VEX}), located at 0.72 AU, our method predicts the ICME arrival speed and time, the magnetic field of the ICME and the resulting \textit{disturbance storm time} (\textit{Dst}) index at Earth. \citet{lin99b} provided pioneering work on the predictability of the \textit{Dst} index with the help of a solar wind monitor located inside 1 AU. We extend this work by introducing a completely new methodology for this type of prediction. So far, typical CME forecasts show an error in arrival time of 6 to 18 hours \citep[e.g.][]{col13,moe14,vrs14,may15,rol16} and cannot provide the magnetic field, or use very general models and relationships to find an approximate \textit{Dst} forecast \citep{tob13,sav15}. Predicting the \textit{Dst} index from data acquired at the Sun--Earth L1 point provides high accuracy but has the disadvantage of a very short prediction lead time of only 30 to 60 minutes \citep[e.g.][]{bal12}. We apply our method to a period where VEX was near the Sun-Earth line that might include contributions from three potentially interacting CMEs, ejected between 2012 June 12--14, that caused a moderate geomagnetic storm with a minimum \textit{Dst} of $-$71 nT on 2012 June 17 14:00 UT.

\begin{figure*}[tbh!]
\epsscale{1.0}
\plotone{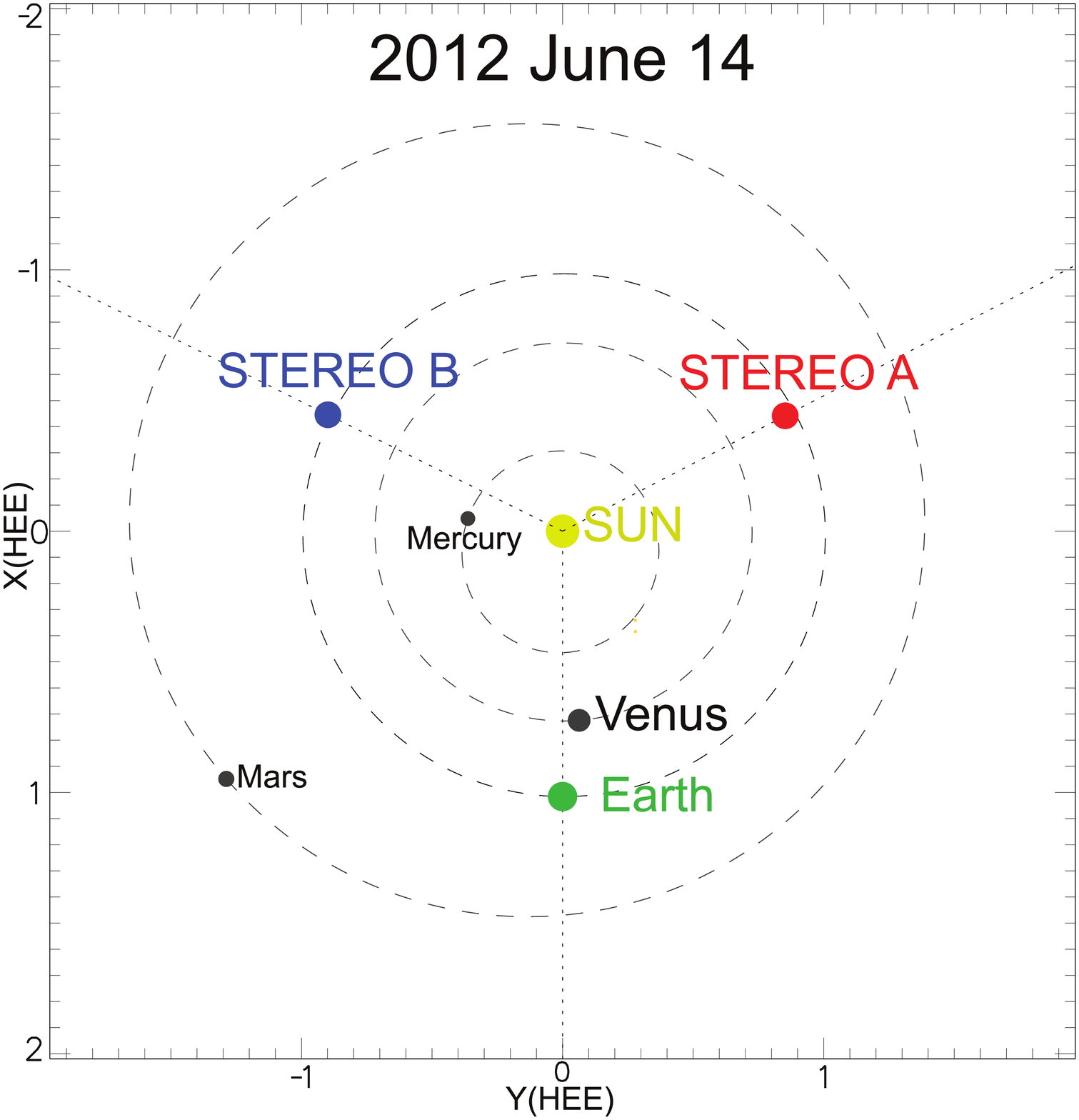}
\caption{Overview of the planetary and spacecraft alignment for the 2012 June 14 CME event. The two \textit{STEREO} spacecraft provide a side view of the CME along the Sun--Earth line. Venus is located $6^\circ$ west of the Sun-Earth line at 0.72 AU.}
\label{fig1}
\end{figure*}

Figure \ref{fig1} depicts the planetary and spacecraft positions on 2012 June 14 with Venus located $6^\circ$ away from the Sun--Earth line, allowing \textit{VEX} data to be used as an in situ reference point close to the Sun--Earth line. The two coronagraphs COR2 \citep[part of the SECCHI instrument suite, ][]{how08} on board the two Solar TErestrial RElations Observatory \citep[STEREO, ][]{kai08} spacecraft provide a stereoscopic side view on the Sun--Earth line and are used to derive CME initial parameters.

In the following sections, we first provide a step-by-step introduction to our method for forecasting \textit{Dst}. We then discuss observations and provide details of three CMEs that preceded the June 17 geomagnetic storm and may have interacted on their way from the Sun to \textit{VEX} and Earth. We also discuss further improvements and applications of our method.

\pagebreak

\pagebreak

\section{Method} \label{sec:method}
We now provide a walkthrough for the necessary steps, parameters and assumptions of our method, starting with the eruption of the CME from the Sun and its evolution through the heliosphere, passing the in situ observing spacecraft and finally leading to geomagnetic activity at Earth. Referring to the constellation Sun -- \VEX -- Earth~L1 (see Figure \ref{fig1}), our method can be summarized in the following steps:

\begin{enumerate}
\item The CME eruption is imaged with coronagraphs, which yields the CME initial speed and direction.
\item The CME propagation is modeled from Sun to \VEX and \WIN with the drag-based model \citep[DBM, ][]{vrs13}. Inputs are the CME initial speed and the solar wind background speed from a WSA-Enlil simulation run  \citep{arg00,ods04}. The output is the CME speed and distance as function of time.
\item The CME propagation with DBM is constrained with the shock arrival time at \textit{VEX}. The input is the time of the CME shock arrival at \textit{VEX}, and the DBM drag-parameter $\gamma$ forms the output of this procedure.
\item The CME magnetic field components observed by \VEX are scaled to Earth L1 in magnitude by using an inverse power law in heliocentric distance derived from previous studies. Then the field is shifted in time according to the kinematics obtained from the previous steps.
\item Finally, using the calculated magnetic field and speed of the CME at Earth L1, the time profile of the \textit{Dst} index is produced.
\end{enumerate}

We now discuss these steps in a more detailed manner before applying them to actual data in Section \ref{sec:data}.

\subsection{CME kinematics} \label{subsec:kinematics}

We use the DBM from \citet{vrs13} to get the time dependent CME position and speed. Strictly speaking, all arrival times and speeds we quote from DBM are valid for the location of the shock at the front of the ICME. The DBM requires three input parameters, which are: the initial speed of the CME, $v_{0}$, at initial distance from the Sun, $r_{0}$, the solar wind background speed, $w$, and the drag parameter, $\gamma$, so that the shock speed, $v(t)$, and distance from the Sun, $r(t)$, as functions of time are given by:

\begin{eqnarray}
v(t) & = & \frac{v_0-w}{1\pm\gamma(v_0-w)t}+w\label{dbm:vt}\\
r(t) & = & \pm\frac{1}{\gamma}\ln[1\pm\gamma(v_0-w)t]+wt+r_0\label{dbm:rt}
\end{eqnarray}

The $\pm$ solution indicates deceleration or acceleration of the CME by the ambient solar wind, where a plus means deceleration ($v_0 > w$) and minus stands for acceleration ($v_0 < w$). 

The CME initial speed is derived from coronagraph observations and for the solar wind background speed $w$ we rely on the WSA-Enlil model because, unfortunately, \textit{VEX} does not provide plasma observations. The background speed is determined by cutting out a solar wind speed profile from WSA--Enlil along the Sun--Earth line from the start of the CME until its shock arrives at the in situ spacecraft. Obtaining $w$ with the help of WSA--Enlil provides a way to avoid using a fixed standard solar wind speed and allows high speed streams to be taken into account. Additionally, this allows an estimate of the speed along the Sun-Earth line that the CME encounters during its propagation from the Sun to Earth, while observations from \textit{VEX}, if available, would only provide the speed at \textit{VEX}.

The drag-parameter, $\gamma$, is a free parameter within a typical range of $0.2 \times 10^{-7}$ to $2 \times 10^{-7} \ \mathrm{km^{-1}} $ \citep{vrs13}, although other studies revealed that smaller values of $\gamma$ on the order of $0.01 \times 10^{-7} \ \mathrm{km^{-1}}$ are also possible \citep{rol14,tem15, rol16}. Knowledge of the shock arrival time at \VEX  allows us to constrain $\gamma$ in a way that the predicted shock arrival from the DBM at \VEX matches the actual, measured arrival time of the shock. This is an important new step, that distinguishes this method from other CME forecasting methods, since once $\gamma$ is fixed, the predicted arrival time and the speed of the shock at Earth L1 can be narrowed down considerably. Using in situ arrival times to constrain CME parameters to study their kinematics has already been introduced by \citet{rol12,rol13,rol14}, but here we apply this concept for the first time in order to predict the CME parameters at another location.

There are some limitations in the use of the DBM with a constant $\gamma$ and solar wind speed $w$ regarding the flexibility to adapt to changes in the ambient medium, for example changes in the solar wind speed or to CME--CME interactions, such as in the event discussed in this paper. For future studies the enhanced DBM, which includes a variability for both $\gamma$ and $w$ as introduced by \citet{zic15} may also be used for describing the CME kinematics within the framework of our method.

\subsection{Dst prediction} \label{subsec:dst}

To forecast the \textit{Dst} index from the predicted CME magnetic field and its speed, we use the two different empirical models \cite[][BU75]{bur75} and \citet[][OB00]{obr00}. Both models require the ICME dynamic pressure, speed and, most important, the magnetic $B_z$  component as input. We can derive the ICME speed from \textit{VEX} observations (see previous Section) and the derivation of $B_z$ will be described in the following paragraphs.

As the CME moves through the heliosphere, its magnetic field strength decreases with increasing distance from the Sun according to a power law, determined from various in situ observations up to $1$~AU and beyond \cite[e.g.][]{bot98, wan05, gul12, win15}. A relation for the magnetic field strength of an ICME at the heliocentric distance between \VEX ($r_{\rm v}$) at 0.72 AU and \textit{Wind} ($r_{\rm w}$) at 0.99 AU was derived by \citet{moe12} using the power law from \cite{lei07}, based on a set of ICME maximum magnetic fields:

\begin{eqnarray}
\label{eqn:b_scaling_mos}
B(r_{\rm w}) & = & B(r_{\rm v}) \left(\frac{r_{\rm w}}{r_{\rm v}}\right)^{(-1.64\pm0.4), }
\end{eqnarray}

with $B(r_v)$ and $B(r_w)$ the magnetic field strength at the position of \VEX and \emph{Wind}, respectively. Applying Equation~(\ref{eqn:b_scaling_mos}) to each of the measured \VEX magnetic field components and the total magnetic field results in an appropriately scaled magnetic field strength at the heliocentric distances of Earth L1. If a CME may become deflected in interplanetary space, which means that it would propagate non-radially away from the Sun, the spacecraft located at Earth L1 could pass through different parts of the CME compared to \emph{VEX}, and the near-Earth spacecraft would measure a deviating magnetic field configuration. However, there is some consensus that the propagation direction of CMEs is largely determined close to the Sun \citep[e.g.][]{kay15,moe15}, so we assume here that no interplanetary deflection is taking place. We are aware that this simple assumption is a limiting case for the accuracy of the method but it enables us to study the potential effects of a deflection of the CME that might happen between the two in situ spacecraft.

We also need to take into account the time it takes for the CME to travel from \VEX to Earth. This requires the results on CME kinematics by the DBM. We shift the \VEX in situ measurements in time by taking the time difference between the measured shock arrival time at \VEX and the predicted arrival time from the DBM at the heliocentric distance of \textit{Wind}.

The last required input parameter for the BU75 and OB00 models is the dynamic pressure. For this parameter, $p_{\rm dyn}$, we assume a typical and constant ICME density of 10 protons cm$^{-3}$ \citep{ric10}. Note that this assumption can lead to an underestimate of $p_{\rm dyn}$ during the ICME sheath region, which often has a higher density, resulting in a first positive peak in \emph{Dst}. However, during the flux rope interval in the ICME the density is low \citep[e.g.][]{ric10}. Thus, the negative excursion in Dst, which is the main phase of the geomagnetic storm that we are mainly interested in predicting, is not expected to be significantly affected by our assumption of constant and low density.

\section{Data} \label{sec:data}

\subsection{Imaging observations} \label{subsec:image_obs}

\begin{figure}[tbh!]
	\epsscale{0.9}
	\plotone{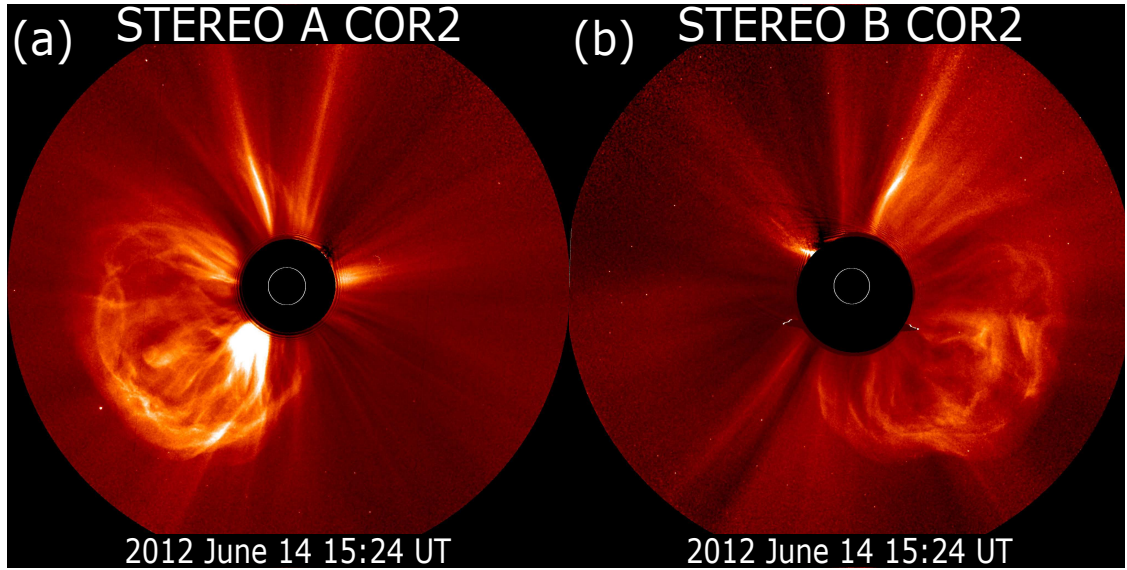}
	\caption{The Earth--directed CME3 on 2012 June 14 observed by the \textit{STEREO-A} (left)  and \textit{STEREO-B} (right) COR2 coronagraphs at 15:24 UT.}
	\label{fig2}
\end{figure}

Figure~\ref{fig2} shows two coronagraph images by the COR2 instrument \citep{how08} on \emph{STEREO-A/B}, about 1 hour after a major Earth-directed CME first appeared on 2012 June 14 14:09 UT. 
However, between 2012 June 12 and 14, there were three potentially Earth directed CMEs, one on each day: CME1 on June 12, CME2 on June 13, CME3 on June 14. In order to understand potential effects of each CME on the subsequently measured in situ data at \VEX and \emph{Wind}, we quote results on interplanetary directions and speeds of these 3 CMEs that were obtained with geometrical modeling based on observations by the \textit{STEREO} Heliospheric Imagers \citep[HI,][]{eyl09}.
We quote values derived from the Fixed--$\Phi$ fitting \citep[FPF,][]{she99,rou08},  self-similar expansion fitting \citep[SSEF, ][with $30$\degree half-width]{dav12,moe13} and harmonic mean fitting \cite[HMF,][]{lug10,moe11} techniques. For a further explanation of these methods see \citet{moe14}. The following results are taken from the HELCATS webpage\footnote{http://www.helcats-fp7.eu/catalogues/wp3\_cat.html, accessed on 2016 July 14.}, a project where catalogs of CMEs are established during the \emph{STEREO} era since April 2007. We quote CME directions in longitude and latitude in the Heliocentric Earth Equatorial (HEEQ) system which are converted from fits to observed time elongation-profiles of the CME leading edges at position angles around the center of the CME \citep[e.g.][]{rou09b,moe14}.

CME1\footnote{HELCATS catalog id: HCME\_A\_\_20120612\_02, HCME\_B\_\_20120612\_01}:
In \textit{STEREO-A/B/COR2} the event appears first on 2012 June 12 18:54 UT, and in \textit{SOHO/LASCO/C3} the event is a partial halo but very faint. Geometrical modeling based on \textit{STEREO-A(B)} yields a direction ranging from $2$\degree to $26$\degree ($-26$\degree to $0$\degreee) in HEEQ longitude and from $-5$\degree to $-7$\degree ($10$\degree to $12$\degreee) in HEEQ latitude, and a speed from 411 to 453 \kmsec (440 to 499 \kmsecc).  Thus, CME1 is an Earth directed CME, with quite consistent results derived from both \emph{STEREO} viewpoints, with a slow interplanetary propagation speed of approximately 450 \kmsecc. 

CME2\footnote{HELCATS catalog id: HCME\_A\_\_20120613\_01, HCME\_B\_\_20120613\_01}: In \textit{STEREO-A/B/COR2} the event appears first around 2012 June 13 14:00  UT. The direction of CME2 ranges from $33$\degree to $38$\degree ($-55$\degree to $-39$\degreee) in HEEQ longitude and from $0$\degree to $1$\degree ($-25$\degreee) in HEEQ latitude. Note that the quoted latitude from \emph{STEREO-A} arises from a tracking at a position angle close to the solar equatorial plane, but in all three coronagraphs (\textit{STEREO-A/B/COR2} and \textit{SOHO/LASCO/C3)}, this slow CME event is strongly southward directed. The speed ranges from $496$ to $500$ \kmsec ($544$ to $574$ \kmsecc).  It is striking that the longitudinal direction is inconsistent comparing the results from the two viewpoints of \textit{STEREO-A/B}, with a difference of about $80$\degreee. In all three coronagraphs the CME also appears asymmetric, which together with the large differences in longitude, may point to a double eruption where neither of the two CMEs is fully earthward directed.

CME3\footnote{HELCATS catalog id: HCME\_A\_\_20120614\_01, HCME\_B\_\_20120614\_01}: The direction ranges from $-6$\degree to $29$\degree ($-36$\degree to $-7$\degreee)  in HEEQ longitude and from $-6$\degree to $-4$\degree ($-1$\degree to $2$\degreee) in latitude with a speed from $791$ to $983$ \kmsec ($877$ to $1017$ \kmsecc). This is a clearly Earth directed (see Figure~\ref{fig2}) and fast CME, with an interplanetary propagation speed of about $900$ \kmsecc. It is unambiguously related to a M1.9 flare at S19E06 peaking at 14:35 UT, close in time to the first image of the CME in COR2A at 14:09 UT. It is accompanied by two dimming regions and a bright post eruption arcade which are coronal signatures interpreted as caused by an erupting flux rope \citep[e.g.][]{web00b}.

In summary, both CME1 and CME3 can be expected to contribute to in situ signatures measured near the Sun - Earth line, and a glancing blow at that location is possible from CME2, due to its mainly southward direction. In order to proceed with calculating the CME kinematics we now need to make an association between an arrival time in the in situ data and the initial speed of one of those 3 CMEs.

\subsection{In situ observations} \label{subsec:insitu_obs}

\begin{figure}[tbh!]
\epsscale{1.0}
\plotone{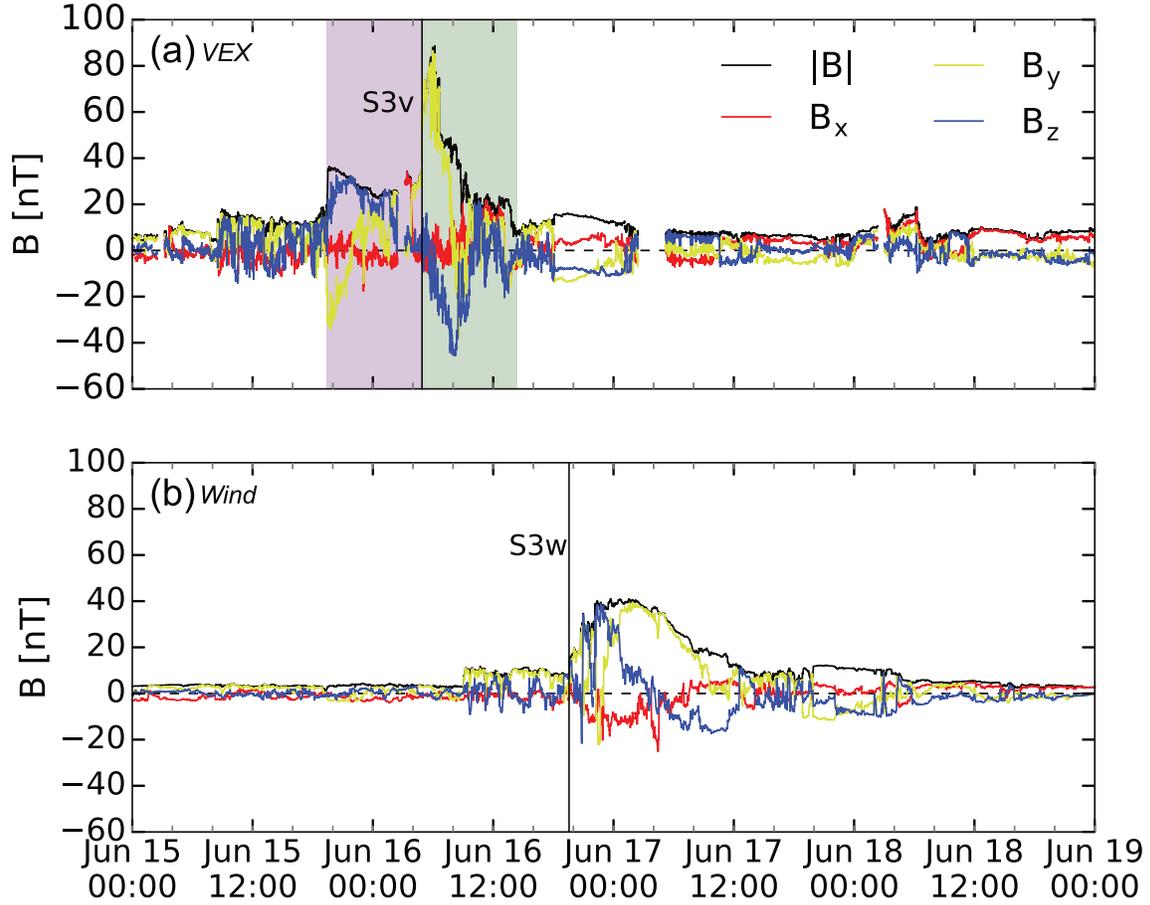}
\caption{In situ magnetic field observations by \textit{VEX} (VSO coordinates) and \WIN (GSE coordinates). (a) Interplanetary magnetic field (IMF) components and total magnetic field at \textit{VEX}. The feature tentatively identified as the shock of CME3, S3v, arriving at \VEX on 2012 June 16 04:54 UT is marked as a vertical black line. Evidence for CME--CME interaction as suggested by the two flux ropes (shaded areas) that immediately follow each other. (b) \WIN observations of the IMF. The feature identified as the shock S3w arrives at \WIN on 2012 June 16 19:34 UT.}
\label{fig3}
\end{figure}

Figure \ref{fig3} shows the in situ magnetic field measurements of \VEX and \emph{Wind}. Since CME3 is Earth directed and has the fastest initial speed of the three CMEs in question, we associate the time where the total field strength makes the most significant jump in the \VEX data as being the shock of CME3 (appropriately named S3v), arriving on 2012 June 16 04:54 UT (vertical solid line in Figure \ref{fig3}a). This association is also based on the observational fact that faster CMEs have a higher in situ magnetic field strength \citep{yur05,moe14}. This means that the high total field following S3v suggests this structure is caused by CME3.
Note that for 3 interacting CMEs this association is very difficult to determine unambiguously, and our association of CME3 with S3v could be challenged by exchanging it with the times of the two other shocks present before S3v. Future studies using our new method will thus need to look in further details concerning this association problem and how it affects the prediction of \emph{Dst}.

Figure \ref{fig3}a also shows that two different flux ropes (two differently shaded areas before and after S3v) immediately follow each other. Magnetic flux ropes are easily discerned in in situ data as they show prolonged, large scale field rotations accompanied by a higher total magnetic field as compared to the background wind \citep[e.g.][]{bur81}. This suggests that at least two of the CMEs merged before arriving at \textit{VEX}, assuming that each CME included a flux rope. The presence of at least two more shocks in the VEX data ahead of S3v possibly generated by CME1 and CME2 is also consistent with merging. The black line in Figure \ref{fig3}b indicates the most likely arrival of shock S3 at the \textit{Wind} spacecraft and is appropriately named S3w, arriving on 2012 June 16 19:34 UT, 14 hours and 40 minutes after arriving at \textit{VEX}. In this study, we do not examine further how the various structures observed in situ at VEX and Wind might be related to the three CMEs. Rather we focus on whether the \textit{VEX} in situ data mapped to Earth can provide a successful prediction of the \emph{Dst} index.

\section{Results} \label{sec:results}

\begin{figure*}[tbh!]
\epsscale{1.1}
\plotone{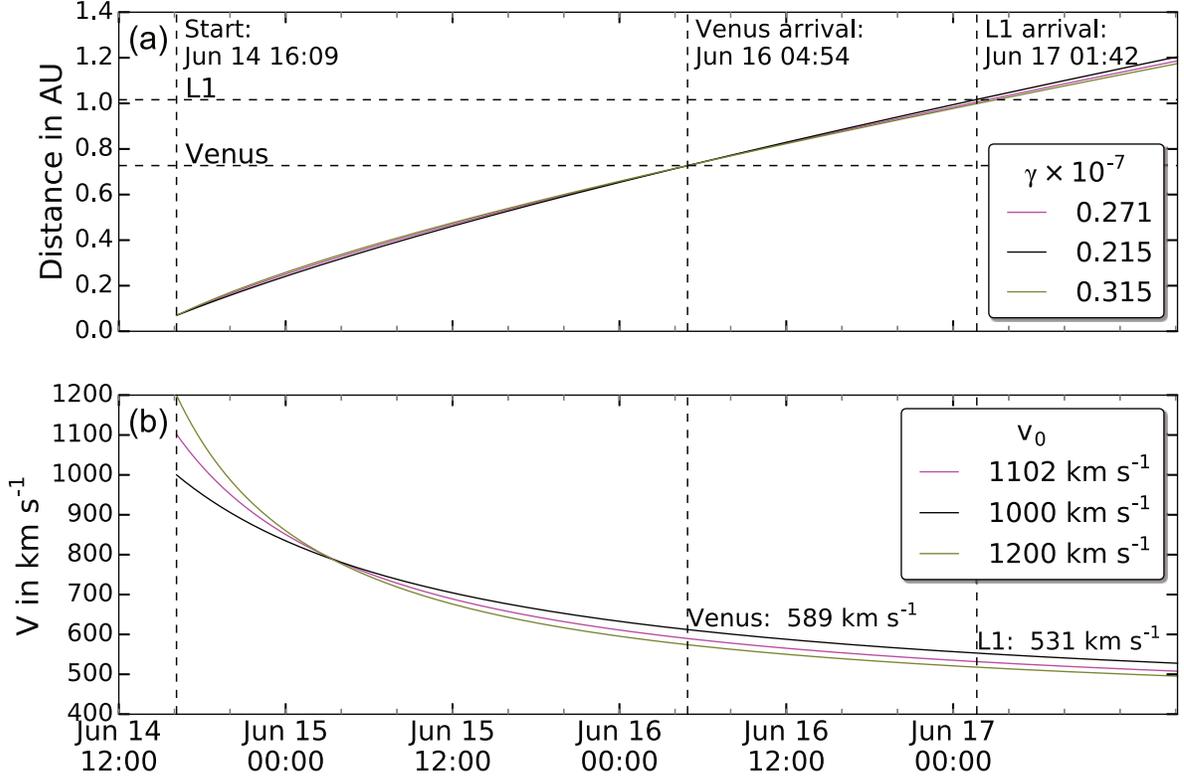}
\caption{CME3 shock kinematics of the 2012 June 14 event, given by the drag-based model (DBM) and constrained with the \VEX shock arrival time, S3v. (a) CME3 shock distance versus time. The horizontal dashed lines mark the distances of \VEX and L1/\textit{Wind}. The vertical dashed lines show the start time of CME3 at 15 solar radii, the arrival time at \VEX (shock time for S3v) and the predicted arrival at L1/\textit{Wind}. (b) CME3 shock speed versus time for the initial speed $v_0$ of 1102 \kmsec (see Section \ref{sec:results}). The predicted speeds for \VEX and \WIN are indicated for $v_0$ = 1102 \kmsec (magenta). Both panels include kinematics arising from variations of $v_0$ to 1200 \kmsec (green) and 1000 \kmsec (black). }
\label{fig4}
\end{figure*}

Figure \ref{fig4} shows the DBM results for $r(t)$ (the heliocentric Distance in AU) and $v(t)$ using an initial speed of CME3 of $v_0 = 1102$~\kmsecc, which was obtained by \citet{moe14} using the graduated cylindrical shell model \citep{the09} at $r_0=15$ solar radii, assuming this speed to be valid for the initial shock speed. We used this speed in favor of the speed mentioned in the data Section because of its higher accuracy. This speed is slightly ($\approx +150$ \kmsecc ) higher than the average interplanetary propagation speed from HI modeling. A range of $v_0$ from 1000 to 1200 \kmsec is taken into account (see black and green lines in the early speed profile in Figure \ref{fig4}b) to include typical uncertainties of speed determination \citep[e.g.][]{moe14}. We have determined the background wind speed by taking an average along the Sun-Earth line in a WSA-Enlil run from the launch of the CME on 2012 June 14 14:09 until the time of the S3v arrival at VEX, resulting in $w=389$ $\pm$ 40 \kmsecc. With the arrival time of the shock S3v at \VEX (2012 June 16 04:54), $\gamma$ is constrained to the value of $0.271\times10^{-7} \ \mathrm{km^{-1}}$ for $v_0= 1102$~\kmsec (Figure \ref{fig4}a). An extrapolation of these kinematics to 0.99 AU in Figure~\ref{fig4}a leads to an arrival time at \WIN on 2012 June 17 01:42. Setting $v_0$ to $1000$ \kmsec and constraining the fit by the VEX arrival time gives a $\gamma$ of $0.215\times10^{-7} \ \mathrm{km^{-1}}$ and a shock arrival time at Wind of 2012 June 17 00:52. Respectively, for a $v_0$ of $1200$ \kmsec we obtain a $\gamma$ of $0.315\times 10^{-7} \ \mathrm{km^{-1}}$ and a shock arrival on 2012 June 17 02:16. This shows, that changing the initial speed from $1000$ \kmsec to $1200$ \kmsec only changes the \textit{Wind} arrival time by 84 minutes in total.

The DBM yields an arrival speed at \textit{Wind} of $531$ \kmsec and an interesting consequence of constraining the DBM fits using the shock arrival time at VEX is that increasing the initial CME speed counter-intuitively moves the arrival time at \textit{Wind} later, and also reduces the arrival speed, for example from $554$ \kmsec for an initial speed of $1000$ \kmsec to $516$ \kmsec for an initial speed of $1200$ \kmsec. The reason for this is that a larger drag coefficient is required for an initially higher speed CME to match the VEX arrival time, and this greater drag then delays the arrival at \textit{Wind} and reduces the arrival speed.

\begin{figure}[tbh!]
\epsscale{0.80}
\plotone{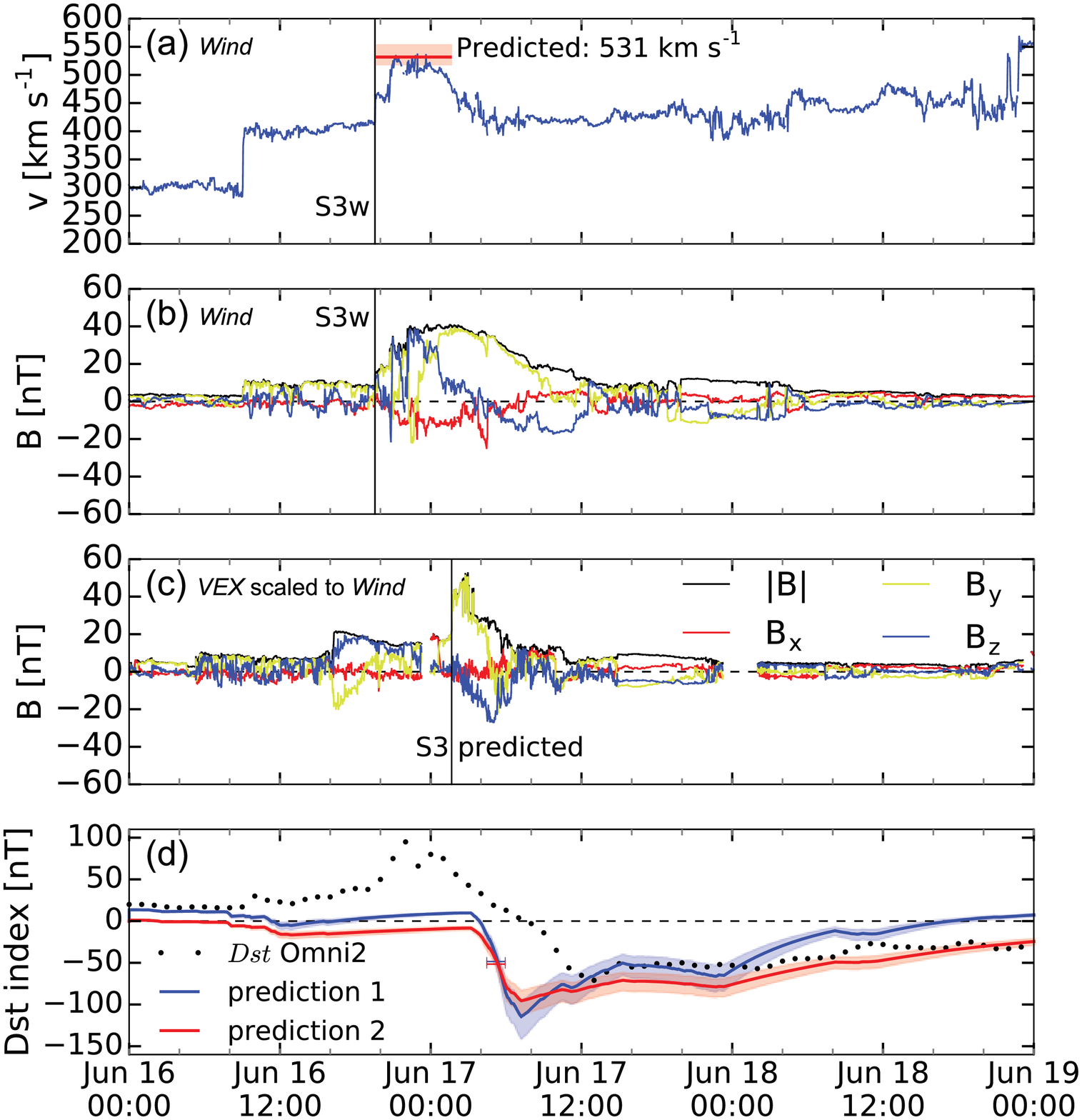}
\caption{
Observations by \WIN compared to predicted magnetic field components based on \emph{VEX} data and geomagnetic response. (a) Solar wind speed measured by \WIN compared to the predicted shock arrival speed and the range in this speed for initial CME speeds of $1000-1200$ \kmsecc, including the arrival time S3w (solid vertical black line). (b) Magnetic field measured by \textit{Wind} with the feature identified as shock S3w marked as solid vertical black line. (c) IMF scaled from \VEX to \WIN by a power law (see Section \ref{subsec:dst}) and time shifted according to the predicted arrival time of the shock S3 at \emph{Wind}, which is indicated as solid vertical black line. (d) Predicted \textit{Dst} using our modeled data as input for \citet{bur75} (blue) and \citet{obr00} (red), compared to the observed \textit{Dst} (black dots). The shaded areas include errors due to uncertainties in CME initial speed and magnetic field scaling factor (see Sections \ref{subsec:kinematics} and \ref{subsec:dst}). The horizontal error bars indicate the errors in the predicted \emph{Dst} profiles due to errors in the Earth arrival time prediction.} 
\label{fig5}
\end{figure}

Figure \ref{fig5}a shows the speed profile observed by the \textit{Wind} SWE instrument \citep{ogi95}, revealing three major jumps. The first jump early on 2012 June 16, to around $400$ \kmsec from a background wind speed of 300 \kmsecc, most likely stems from one of the previous CMEs. The second jump in speed, labeled S3w, is followed by another jump shortly afterwards. This double--shock feature could stem from two merged CMEs and makes the association of a shock to its corresponding CME somewhat ambiguous, therefore limiting the accuracy in arrival time prediction. Figure \ref{fig5}a also shows the predicted arrival speed (red horizontal line) and its range ($516-554$ \kmsecc, shaded red area) based on initial CME speeds of $1000-1200$ \kmsecc. This is consistent with the observed solar wind speeds following the third shock. The predicted speed of $531 \pm 23$~\kmsec (Figure \ref{fig5}a) only deviated by 41~\kmsec from the measured value of $490 \pm 30$~\kmsecc, for which we took an average speed during the sheath region, from S3w to June 17 00:00 UT.

For calculating the \emph{Dst} index, the ICME speed was set to 531 $\pm$ 23 \kmsec (including the variations of $v_0$ as described above) throughout the ICME and $B_z$ is taken from the scaled and time shifted \VEX observations, done according to Section~\ref{subsec:dst}. We shift the \VEX in situ measurements in time by taking the time difference of 20 hours and 46 minutes between predicted arrival time from the DBM at L1 (June 17 01:40) and the measured shock arrival time of CME3 (S3v) at \VEX (June 16 04:54). This time of about 21 hours is also the lead time of the prediction. Figure \ref{fig5}c shows a prediction of the magnetic field corrected for both strength and timing at Earth L1 by applying our method to measurements from \VEX at 0.72 AU. Figure \ref{fig5}b shows the magnetometer data from \textit{Wind} for a direct comparison, demonstrating a decent though certainly not perfect match in timing, field strength and the behavior of $B_z$.

Figure \ref{fig5}d shows the prediction of the \textit{Dst} index with our method, solely based on scaled data from 0.72 AU, in comparison to the actual \textit{Dst} (black dots, provided by the World Data Center for Geomagnetism, Kyoto). The blue line is the \textit{Dst} prediction using the BU75 model and red shows the OB00 result. Shaded areas mark vertical errors in \textit{Dst}, which arise (1) from the uncertainty in the exponent of the power law in Equation (\ref{eqn:b_scaling_mos}) and (2) from a variation of the predicted CME arrival speed at Earth L1. The horizontal error bars in Figure~\ref{fig5}d indicate errors in the predicted arrival time due to uncertainties in the ICMEs initial speed $v_0$ $\pm$ 100 \kmsecc. For better visibility, these horizontal bars are only drawn for the period of decreasing \textit{Dst}. The measured \textit{Dst} shows a minimum of $-71$~nT and the BU75 (OB00) models yield a minimum \textit{Dst} of $-96$ ($-114$)~nT with an upper limit of $-83$ ($-96$)~nT and a lower limit of $-113$ ($-141$)~nT. The average predicted values minimum \emph{Dst} values are $-25$ nT (BU75) and $-43$ nT (OB00) below the observed one, which is a rather promising result.

\section{Discussion} \label{sec:discussion}
To calculate a prediction for the \textit{Dst} index out of measurements from \textit{VEX} at 0.72 AU, we started with the DBM  to obtain the arrival time and the speed of the CME at L1. The solar wind background speed for the DBM was derived by data from the WSA-Enlil model rather than setting the solar wind speed to a default value. In doing so, we are able to take solar wind high speed streams into account. There is potential for improving the method regarding cases where a CME is traveling through more distorted solar wind.

Note that, counterintuitively, the highest $v_0$ results in the latest arrival time, as well as the lowest arrival speed at \textit{Wind} and vice versa, because $\gamma$ is constrained to the ICME arrival time at the in situ spacecraft (see also Figure~\ref{fig4}). Determination of $\gamma$ in this way has the advantage of a good estimate of its average value up to the in situ spacecraft. On the other hand, constraining $\gamma$ at the in situ spacecraft and assuming a constant $\gamma$ all the way from the Sun to Earth may lead to inaccurate results if the in situ spacecraft (the one at $< 1$~AU) would be located at a heliocentric distance closer to the Sun, and a variable $\gamma$ might be taken into account \citep{zic15}.

Compared to the true shock arrival at L1 on 2012 June 16 19:34 UT (S3w, Figure \ref{fig5}b) the DBM predicted the shock at 2012 June 17 01:40 UT (S3, Fig. \ref{fig5}c), about 6 hours too late, which is clearly above the error of $\pm$ 1 hour associated with this arrival time from our calculations. As an explanation for the discrepancy between predicted and actual arrival time of 6 hours we suggest two hypothesis: (1) A local distortion of the shock front. (2) The association between the measured magnetic field structures that arrive at \VEX and \WIN is ambiguous for this event and might be incorrect. 

For point (1), consider the following: \VEX is located about 6\degree away from the Sun--Earth line in heliospheric longitude which is equivalent to 0.075 AU (at \VEX distance) and 0.105 AU (at Earth distance). An interaction of CME3 with one or both of the two preceding CMEs might have distorted the CME shock shape in such a way that the part of shock S3 that travels along the Sun--Earth line has already traveled further in heliocentric distance than the part that impacted \textit{VEX}. This could be explained by an inclination of the shock surface. However, if we assume that the shock traveled for 6 hours with a speed of 530 \kmsecc, the resulting radial distance between the the part of the shock that hits \VEX and the part that travels along the Sun--Earth line would be in the order of 0.08 AU, thus the possible inclination of the shock surface would be very significant. If hypothesis (2) holds and the magnetic field structures at \VEX and \WIN are incorrectly associated, the error between predicted and measured arrival could be different, but this ``association problem'' will be further discussed in future studies. 

Aside these two hypotheses, we can rule out one other potential factor that could be responsible for the arrival time discrepancy, namely an incorrect determination of the DBM drag parameter $\gamma$. In order to cover the distance of 0.28 AU from \VEX to Earth L1 within the time interval 2012 June 16 04:54 (S3 at \textit{VEX}) and June 16 19:34 (S3w at \textit{Wind}) it would require an average speed of $\sim$790~\kmsec. This means that the speed at \VEX would even be way higher, leading to an unlikely scenario where virtually no deceleration of the shock takes place from the Sun to \VEX (low $\gamma$) and all the deceleration happens between \VEX and Earth (very high $\gamma$). Such a scenario might be possible if CME3 had interacted with one of the preceding CMEs somewhere between \VEX and Earth, but \VEX data shows that the interaction already occurred before CME3 hit \VEX (see Figure \ref{fig3}a).

To predict the magnetic field at Earth L1, we have applied Equation (\ref{eqn:b_scaling_mos}) to each magnetic field component from \VEX measurements individually, although previous work on scaling of the magnetic field inside flux ropes suggest that the axial and azimuthal components may scale with $1/r^2$ and $1/r$ respectively \citep[e.g.][]{osh93, far93}. However, the exponent $-1.64 \pm 0.4$ covers most of this range and a different scaling law may be added in future studies.

\section{Conclusions} \label{sec:summary}
We showed that the problem of predicting the geomagnetic effects of CMEs may in principle be solved by using in situ magnetometer observations by a spacecraft located in the inner heliosphere close to the Sun-Earth line, thereby strongly extending the method of \cite{lin99b}. The position of the in situ spacecraft is a trade-off between prediction accuracy and lead time, where a spacecraft close to Earth provides high accuracy forecasts at the cost of short lead time. We have studied a CME which led to a moderate geomagnetic storm at Earth and likely interacted with two preceding CMEs. We applied for the first time a new method consisting of several steps to predict the \textit{Dst} time profile. With data from \textit{VEX}, located at $0.72$, AU we obtained a relatively accurate \textit{Dst} forecast ranging from of $-83$ to $-141$ nT (including error bars), near the observed \textit{Dst} of $-71$ nT, while still maintaining a prediction lead time of nearly $21$ hours. Knowing the time of CME ejection at the Sun on 2012 June 14 14:09 UT and the arrival at \VEX on 2012 June 16 allows us to tightly constrain the CME kinematics, leading to a predicted CME arrival speed at Earth of $531 \pm 23$~\kmsecc, similar to the observed $490 \pm 30$~\kmsec at \textit{Wind}.

The predicted arrival time of the CME at Earth is 2012 June 17 01:40 UT, which is $6 \pm 1$ hours later than the true shock arrival. This discrepancy in arrival time might be explained by an incorrect association between the measured magnetic field structures at \VEX and \WIN or by an inclination of the shock surface on a scale of 0.08 AU in radial distance. If the shock surface indeed shows such a pronounced local feature, this means that there may exist a limit for the errors in arrival time prediction of CMEs on the order of 6 hours, which cannot be further improved unless the shock shape is modeled very accurately on scales of $< 0.1$~AU. Our method thus also provides a new tool for detailed diagnostics on CME physics that allows to reveal those small-scale structures.

Our obtained results for \textit{Dst} are relatively accurate despite the interaction of several CMEs before they impacted \textit{VEX}  and despite the assumptions that were made regarding the solar wind background speed estimation, magnetic field scaling and determination of the drag parameter $\gamma$. To further improve the accuracy of the results, a better solar wind background speed estimation can be made by a plasma instrument on the in situ spacecraft. This would allow to map a speed profile that includes any variations in the solar wind background speed, and it would allow an even more constrained calculation of the CME kinematics.

The DBM we used is designed for constant input parameters, so further studies on more events and different heliospheric distances of the in situ spacecraft will reveal if the assumption of a constant $\gamma$ and $w$ is sufficient, or if the enhanced DBM \citep{zic15} with variable $\gamma$ and $w$ may provide better results.

Other improvements of our method include modeling of the expansion of the CME flux rope \citep[e.g.][]{dem09}, even with different power laws for the azimuthal and axial fields \citep[e.g.][]{osh93}, a conversion of the predicted $B_z$ component to GSM coordinates, including the CME shock shape \citep{moe13,moe15}, adding a variable dynamic pressure profile, and using an updated method for predicting \emph{Dst} from solar wind parameters \citep{tem06}. Because we have assumed a constant ICME density, the positive peak in \textit{Dst} in Figure \ref{fig5}d, which stems from a very high density and thus dynamic pressure, is not included in the \textit{Dst} forecast. However, this peak maximizes at +95 nT late on 2012 June 16, which is actually the highest positive \textit{Dst} value in the entire space age, at least since 1957. The impact of these CMEs on Earth has resulted in a strong magnetospheric compression, and a study of the exact interplanetary causes would be worthwhile.

An application of our method to real time magnetic field data by possible future spacecraft thus provides the possibility to make appropriate and timely arrangements to prepare for the potential effects of CMEs on the technological infrastructure of mankind. Our proof-of-concept study may be extended, using any spacecraft temporarily located close to the Sun-Earth line \citep[see][]{lin99b}, such as \textit{MESSENGER} in conjunction with an L1 spacecraft. In addition any line-up of two spacecraft like \textit{VEX} -- \textit{STEREO} or \textit{MESSENGER} -- \textit{STEREO} can be used to predict CME arrival times and speeds, as well as magnetic fields from the inner to the outer spacecraft \citep[e.g.][]{goo15, win16}. It can be applied to future space missions in the inner heliosphere like \textit{BepiColombo}, \textit{Solar Probe Plus} or \textit{Solar Orbiter}, or to data from permanent in situ observers at an artificial Lagrange point like the \textit{Sunjammer} concept \citep{eas15}, which would allow a continuous, gap-free forecast of the \textit{Dst} index. It may also be applied to data by a future ring of spacecraft around the Sun  \citep{rit15}, carrying at least a magnetometer, where each spacecraft would drift towards and away from the Sun--Earth line. Its feasibility and number of spacecraft needed depends mainly on the longitudinal extent of CME flux ropes \citep[e.g.][]{bot98,goo16}, which also warrants further study.

\acknowledgments

M.K, C.M. and T.A. thank the Austrian Science Fund (FWF): [P26174-N27]. The presented work has received funding from the European Union Seventh Framework Programme (FP7/2007-2013) under grant agreement no. 606692  [HELCATS].

\newpage

\bibliographystyle{apj}

\end{document}